\documentclass{moriond}

\usepackage{amssymb}
\usepackage{amsmath}

\DeclareMathOperator{\Imm}{Im}

\def\be{\begin{equation}}
\def\ee{\end{equation}}
\def\bea{\begin{eqnarray}}
\def\eea{\end{eqnarray}}

\newcommand{\Photo}{}

\begin{document}


\title{54TH RENCONTRES DE MORIOND:
\\QCD AND HIGH ENERGY INTERACTIONS\\ \vspace*{1cm} TRANSPORT ANOMALIES NEAR THE QCD PHASE TRANSITION AT FINITE DENSITY}

\author{B.O. Kerbikov\vspace*{0.8cm}}
\address{A.I.~Alikhanov Institute for Theoretical and Experimental Physics,
Moscow 117218, Russia}
\address{Lebedev Physical Institute,
Moscow 119991, Russia}
\address{Moscow Institute of Physics and Technology, Dolgoprudny 141700, Moscow Region, Russia}

\maketitle\abstracts{
We show that at finite density pressure-pressure and current-current correlators exhibit divergences at $T_c$ owing to the fluctuations of the diquark field. Specifically, this leads to a significant excess of the soft photon production rate near $T_c$.
} 

A large body of experimental data on heavy ion collisions obtained at RHIC and LHC has lead to a revolutionary change in our view on the properties of QCD matter at finite temperature and density. These properties depend on the location of the system in the QCD phase diagram, i.e. on the values of the temperature and the chemical potential. At present the most intriguing is the regime of finite density and low or moderate temperature. On the theoretical side we understand much better what happens to quark-gluon matter at high temperature and zero or small $\mu$ since this domain of the phase diagram is accessible to lattice calculations. At nonzero density one has to resort to effective theories or models like NJL. At this point necessary to mention the new emerging approach to quark-gluon thermodynamic at finite density \cite{01}.

Our focus in the present talk is on the finite density pre-critical fluctuation region with $T \to T_c$ from above. Comprehensive study has shown that at high density and low temperature the ground state of QCD is color superconductor \cite{02,03}. We consider the 2SC color superconducting phase when $u$- and $d$- quarks participate in pairing but the density is not high enough to involve the heavier $s$- quark in pairing. The value of the quark chemical potential under consideration is $\mu \simeq 200$-$300$ MeV and the critical temperature $T_c \simeq 40$ MeV. The corresponding density is two to three times the normal nuclear matter density. Both numbers should be considered as an educated guess since they rely on model calculations.

An important difference of color superconductor from the BCS one is that instead of an almost sharp border in BCS between the normal and superconducting phases the transition in color superconductor is significantly smeared \cite{04}. The fluctuation contribution to the physical quantities is characterized by the Ginzburg-Levanyuk number $Gi$ which for the quark matter can be estimated as \cite{04}
\be
Gi \simeq \frac{\delta T}{T_c} \simeq \left( \frac{T_c}{\mu}\right)^{4} \lesssim 10^{-3}, 
\label{01}
\ee 
where $\delta T$ is the width of the fluctuation region. Note that for BCS superconductor $Gi\sim 10^{-12}$-$10^{-14}$. 

We want to investigate the pressure and the electromagnetic response of the above fluctuation state, i.e. the temperature dependence of the energy-momentum and current-current correlators. It is known that these correlators, or the related response operators, can be evaluated only in perturbation theory. The related physical observables are: (i) bulk viscosity and sound attenuation, (ii) electrical conductivity and soft photon emissivity. It will be conjectured that these observables diverge at $T \to T_c$ as $(T-T_c)^{-3/2}$ for (i) and as $(T-T_c)^{-1/2}$ for (ii).

\begin{figure}
\begin{minipage}{0.99\linewidth}
\centerline{\includegraphics[width=0.35\linewidth]{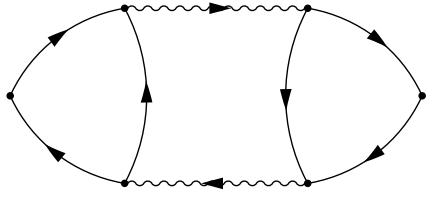}}
\end{minipage}
\caption[]{Feynman diagram for the AL polarization operator.}
\label{fig:FIG1}
\end{figure}

The dynamical origin of fluctuations is the soft mode of the diquark field. In the vicinity of $T_c$ non-equilibrium quark pairs are formed with the characteristic Ginzburg-Landau life time $\tau_{GL} \sim (T-T_c)^{-1}$. Precursor pair fluctuations above $T_c$ give the dominant contribution to the quark transport coefficients. The leading diagram defining the retarded response operator is the Aslamazov-Larkin (AL) \cite{05,06} one shown in Fig.1. It includes two singular at $T_c$ fluctuation propagators (FP) of the diquark field depicted by wavy lines. The solid lines are quark Matsubara propagators, the in- and out- vertices stand for quark-phonon and quark-photon interactions. Quark pairs formation like Cooper pairs formation is a non-perturbative process. In this sense the AL diagram is a non-perturbative one. 

Relativistic FP has been evaluated using either Dyson equation \cite{07}, or time-dependent Landau-Ginzburg equation with stochastic Langevin forces \cite{08}. It reads
\be
L({\mathbf k},\omega)=-\frac{1}{\nu}\,\frac{1}{\varepsilon + \frac{\pi}{8T_c}(-i\omega+D{\mathbf k}^2)}.
\label{02}
\ee
Here $\nu=\frac{p_0 \mu}{2 \pi^2}$ is relativistic density of states at the Fermi surface with $p_0$ being the Fermi momentum, $\varepsilon = (T-T_c)T_c^{-1}$, $D$ is the diffusion coefficient. The two FP-s entering into the AL diagram lead to the divergence of the transport coefficients at $T \to T_c$. 

The finite-temperature retarded response function can be symbolically written as \cite{07,08}
\be
\Pi = -4 Q^2 T \sum \int BLBL,
\label{03}
\ee
where summation goes over the internal Matsubara frequencies and integration is over the internal momenta. The coupling is $Q^2=\frac59 e^2$ for the electromagnetic mode with two flavors and $Q^2=g^2$ for the sound mode. The FP $L$ is defined by (\ref{02}), $B \sim GGG$ is the block of three Matsubara propagators shown in Fig.1. Different character of coupling of the two modes (vector and scalar) induces important difference into the factros $B$. To get nonzero $B_s$ for the sound mode one has to take into account the energy dependence of the density of states at the Fermi surface and to introduce the ultraviolet cutoff $\Lambda$ so that $B_s \sim log (\Lambda / 2\pi T_c)$ \cite{05,08,09}. Keeping only linear in the external frequency $\omega$ terms one arrives at  
\be
\Pi_{em}=-i\omega\,\frac{B_{em}}{12\nu^2}\int\frac{d\mathbf{q}}{(2\pi)^3}\frac{\mathbf{q}^2}{(\varepsilon + \frac{\pi}{8T_c}D\mathbf{q}^2)^3}
\label{04}
\ee
\be
\Pi_{s}=-i\omega\,\frac{B_{s}}{\nu^2}\int\frac{d\mathbf{q}}{(2\pi)^3}\frac{1}{(\varepsilon + \frac{\pi}{8T_c}D\mathbf{q}^2)^3}
\label{05}
\ee
For the electrical conductivity $\sigma$ and the sound attenuation coefficient $\gamma$ this yields
\be
\sigma = -\frac{1}{\omega}\Imm\Pi_{em} = \frac{e^2}{16}\left( \frac{\pi D}{8T_c} \right)^{-1/2}\left( \frac{T-T_c}{T_c} \right)^{-1/2},
\label{06}
\ee
\be
\gamma = \omega^2 g^2 A \log^2\frac{\Lambda}{2\pi T_c}\left( \frac{T-T_c}{T_c} \right)^{-3/2},
\label{07}
\ee
where $A=m^2(2p_0)^{-4}\kappa^{-3}$, $\kappa^2=\frac{\pi D}{8T_c}$. The sound attenuation per wavelength is $\alpha = \gamma \lambda \sim \omega \varepsilon^{-3/2}$. The rise of the acoustic attenuation near $T_c$ results in strongly divergent bulk viscosity $\zeta(T) \sim \varepsilon^{-3/2}$. This temperature dependence is rather close to the scaling law $\zeta \sim \xi^{z - \alpha/\nu}$ \cite{10,11}, where $\xi$ is the correlation length, $z \simeq 3$ is the dynamical critical exponent, $\nu \simeq 0.6$ is the correlation length critical exponent, $\alpha \simeq 0.11$ is the critical exponent of the heat capacity. 

Along with the electromagnetic conductivity the current-current correlator gives rise to the photon emissivity which is expressed through the imaginary part of the retarded photon self-energy as \cite{12}
\be
\omega \frac{d R}{d^3 p}=-\frac{2}{(2\pi)^3}\Imm\Pi_{em}\frac{1}{e^{\omega /T}-1},
\label{08}
\ee
where $\Pi_{em}$ is given by the diagram shown in Fig.1. This diagram defines the electrical conductivity and the photon emissivity to the order $\alpha$ in the electromagnetic sector and non-perturbatively in strong interactions since pair formation in the vicinity of the Fermi surface is not calculable in perturbation theory. Comparing (\ref{06}) and (\ref{08}) we find
\be
\lim\limits_{\omega \to 0}\omega \frac{d R}{d^3 p}=\frac{1}{4\pi^3} T \sigma.
\label{09}
\ee 
As an illustration we present a numerical estimate of the photon emissivity. We take $T_c=40$ MeV and $Gi \simeq \varepsilon = 10^{-3}$. Note that the linear fluctuation theory breaks down at some small value of $\varepsilon$ which is difficult to estimate. We also need the diffusion coefficient equal to $D=\frac{1}{3} v^2 \tau \simeq 0.17$ fm under the assumption $v=1$, $\tau = 0.5$ fm.

With the above set of parameters we obtain
\be
\sigma\simeq 0.18 \text{ fm}^{-1},\quad \lim\limits_{\omega \to 0}\omega \frac{d R}{d^3 p} \simeq 0.73 \cdot 10^{-2} \text{ fm}^{-4}\text{ GeV}^{-2}.
\label{10}
\ee
According to (\ref{06}) and (\ref{09}) the soft photon production is enhanced in the vicinity of $T_c$ which may be a tentative suggestion for the FAIR/NICA investigation.

\section*{Acknowledgments}

This work has been supported by a grant from the RFBR number 18-02-40054. The author's participation in Rencontres de Moriond QCD 2019 Conference was a part of MIPT ``5-100'' Program.

\end{document}